\documentclass[twocolumn,showpacs,preprintnumbers,amsmath,amssymb,pra]{revtex4}

\usepackage{graphicx}

\newcommand{\ket}[1]{\vert #1\rangle}
\newcommand{\bra}[1]{\langle #1\vert}

\newcommand{\ii}{\mathrm{i}}

\newcommand{\diff}{\mathrm{d}}
\newcommand{\eh}{\mathrm{e}}

\newcommand{\up}{\uparrow}
\newcommand{\down}{\downarrow}
\renewcommand{\Re}{\mathfrak{Re}}

\DeclareMathOperator{\teopnull}{\hat{\mathcal{U}}_0}
\DeclareMathOperator{\teop}{\hat{\mathcal{U}}}

\begin{document}

\title{Simulating high-temperature superconductivity model Hamiltonians
with atoms in optical lattices}

\author{Alexander Klein and Dieter Jaksch}

\affiliation{Clarendon Laboratory, University of Oxford, Parks Road,
Oxford OX1 3PU, United Kingdom}

\begin{abstract}
  We investigate the feasibility of simulating different model
  Hamiltonians used in high-temperature superconductivity.
  We briefly
  discuss the most common models and then focus on the simulation of
  the so-called $t$-$J$-$U$ Hamiltonian using ultra-cold atoms in
  optical lattices. For this purpose, previous simulation schemes to
  realize the spin interaction term $J$ are extended. We especially
  overcome the condition of a filling factor of exactly one, which
  otherwise would restrict the phase of the simulated system to a
  Mott-insulator. Using ultra-cold atoms in
  optical lattices allows simulation of the discussed models for a very wide
  range of parameters.  The time needed to simulate the Hamiltonian is
  estimated and the accuracy of the simulation process is numerically
  investigated for small systems.
\end{abstract}
\date{\today}
\pacs{03.75.Ss, 03.67.-a}

\maketitle

\section{Introduction}

Since its discovery, high-temperature superconductivity
\cite{Bednorz-ZPhysB-1986} has attracted much attention on the
theoretical as well as on the experimental side. Nevertheless, the
mechanisms which lead to this effect are not completely understood
\cite{Basov-RMP-2005}. On the theoretical side one encounters the
problem that the physics of a very complex many-body system has to
be described and calculated. Because quantum effects play a crucial
role, the calculation of these systems is a very hard task on a
classical computer. Furthermore it is not certain that the
investigated theoretical models accurately describe the macroscopic
features of superconducting materials used in experiments.
Experimental efforts to test these models have suffered from
difficulties in changing the system parameters for a given
superconductor. In general different parameter sets require
different types of superconducting material. Due to all these
obstacles it is worthwhile searching for a versatile system which
can be used to accurately simulate high-temperature superconductor
models.

As we will show in this paper ultra-cold atoms in optical lattices
are a very good candidate for simulating high-temperature
superconductivity. During the past few years the field of cold atom
physics has made tremendous progress and entered the regime of
accurately controlled strongly correlated systems. Seminal
experiments with bosonic atoms have, e.g.,~realized the
Mott-insulator to superfluid transition and cold controlled
collisions between atoms in optical lattices
\cite{Greiner-Nature-2002, Greiner-Nature-2002b, Mandel-Nature-2003,
Mandel-PRL-2003}. They found excellent agreement with corresponding
theoretical models \cite{Jaksch-PRL-1998, Jaksch-PRL-1999}. For
clouds of fermionic atoms superfluidity has been demonstrated by
creating vortices \cite{Zwierlein-Nature-2005} and Fermi surfaces of
atoms loaded into three-dimensional optical lattices have been
measured \cite{Koehl-PRL-2005}. Optical lattice systems are also
very flexible; a large range of Hamiltonians can be realized and
system parameters are easily varied over a wide range by changing
external laser parameters \cite{Jaksch-Ann-2005}. For instance, it
has also been proposed to implement effective magnetic fields
\cite{Jaksch-NJP-2003, Juzeliunas-PRL-2004, Juzeliunas-PRA-2005,
Juzeliunas-PRA-2006, Mueller-PRA-2004, Sorensen-PRL-2005} and even
the implementation of non-Abelian gauge fields seems to be
feasible~\cite{Osterloh-PRL-2005, Ruseckas-PRL-2005}.

We will briefly discuss some of the model Hamiltonians which are
used to describe high-temperature superconductive cuprates. We will
then focus on the $t$-$J$-$U$ Hamiltonian \cite{Zhang-PRL-2003},
which includes a spin-spin coupling term $J$ additionally to hopping
$t$ and on-site interaction $U$. The simulation of this Hamiltonian
can be used for obtaining a deeper understanding of the
corresponding model, and also enlarges the parameter space by an
additional, freely tunable interaction term $J$. In analogy to the
cuprates we expect this model to show a very rich phase diagram for
the atoms in the optical lattice including the appearance of
(anti)ferromagnetic phases.

The simulation of the spin-spin interaction part in optical lattices
has already been discussed in Refs.~\cite{Jane-QIC-2003,
Sorensen-PRL-1999, Duan-PRL-2003}. Atoms with two internal states
simulate a spin chain and the spin-spin interaction is implemented
by state dependent shifting of the atoms causing controlled atomic
collisions. These collisions induce state dependent phase shifts of
the atomic wave function mimicking the spin-spin interaction.
However, all of the proposals \cite{Jane-QIC-2003,
Sorensen-PRL-1999, Duan-PRL-2003} require a filling of very close to
or exactly one atom per lattice site. For superconductors, this
corresponds to a Mott-insulating state. In order to extend this
method to the simulation of the $t$-$J$-$U$ Hamiltonian we will
relax the condition of having exactly one atom in each lattice site.
Additional shifts of the atoms will be necessary for simulating the
spin interaction term $J$ and we will also show that by choosing
appropriate parameters it is possible to implement effectively
attractive interaction terms $U<0$.

The simulation of the time evolution of the whole $t$-$J$-$U$
Hamiltonian is performed via a Trotter-Suzuki expansion
\cite{Trotter-PAMS-1959, Suzuki-PLA-1990, Suzuki-JMP-1991,
Suzuki-JMP-1993, Suzuki-CMP-1994, Jane-QIC-2003} and we will
consider the case of temperature $T=0$ in our calculations. To be
able to experimentally observe superconductivity effects using our
simulation method the temperature of the optical lattice atoms will
have to be lower than $k_\mathrm{B} T<0.02 t$
\cite{Dagotto-RMP-1994}, where $k_\mathrm{B}$ is the Boltzmann
constant. Such low temperatures are experimentally difficult to
achieve. However, theoretical proposals for fault tolerant loading
of fermionic atoms \cite{Griessner-PRA-2005} into the lowest
motional band of an optical lattice and phonon assisted side band
cooling within this motional band \cite{Griessner-PrivCom-2005}
exist. These methods will be realizable with current and near future
experimental techniques and enable achievement and control of
temperatures sufficiently small for our purpose.

This work is organized as follows. In Sec.~\ref{Sec:ModHams} we
briefly discuss some of the model Hamiltonians used to describe
high-temperature superconductivity in cuprates. In
Sec.~\ref{Sec:Sim_in_latt} we explain how one of these Hamiltonians,
namely the $t$-$J$-$U$ model, can be simulated using ultra-cold
atoms in optical lattices. The simulation process is numerically
tested in Sec.~\ref{Sec:Numerics} and we conclude in
Sec.~\ref{Sec:Concl}.

\section{Model Hamiltonians for high-$T_c$ superconductors
\label{Sec:ModHams}}

In this section we briefly discuss some of the model Hamiltonians
which are used to describe high-temperature superconductors
\cite{Hirsch-PRL-1985, Zhang-PRB-1988,Anderson-Science-1987}. The
most basic microscopic description is given by the Hubbard model
\begin{equation}\label{Eq:HtU}
  \hat H_{tU} = - t \sum_{\langle i,j \rangle, \sigma}  \hat
  c^\dagger_{i,\sigma} \hat c_{j,\sigma}  + U \sum_j
  \hat n_{j,\up} \hat n_{j,\down} \,.
\end{equation}
The operator $\hat c_{j,\sigma}^\dagger$ ($\hat c_{j,\sigma}$)
creates (annihilates) an electron with spin $\sigma = \up, \down $
in lattice site $j$ and obeys the standard fermionic anticommutation
relations, the number operator is denoted by $\hat n_{j,\sigma} =
\hat c^\dagger_{j,\sigma} \hat c_{j,\sigma}$. This Hamiltonian
describes electrons in the lowest Bloch band tightly bound to the
lattice sites formed by the atoms of the solid. The electrons can
tunnel from one lattice site to the nearest-neighbor site (indicated
by the brackets $\langle i,j \rangle$) with hopping energy $t$ and
the repulsive on-site Coulomb interaction is given by $U$. Because
of the Pauli exclusion principle for electrons and the single Bloch
band assumption in Hamiltonian Eq.~(\ref{Eq:HtU}), there can be a
maximum of one electron of each spin state in a single lattice site.
Hence no onsite interaction term for electrons in the same spin
state is included in the Hamiltonian.

The Hamiltonian Eq.~(\ref{Eq:HtU}) describes the behavior of the
electron gas in two-dimensional (2D) layers. If the lattice sites
are all occupied by a single electron the system behaves like a
Mott-insulating anti-ferromagnet. By hole-doping of the cuprate the
behavior of the gas changes considerably and for a critical doping
and temperatures below $T_c$ the solid gets superconductive (see,
e.g.,~\cite{Dagotto-RMP-1994,Damascelli-RMP-2003}). In a real
superconductor many of these two-dimensional layers are stacked on
top of each other. The influence of these layers can be included by
a small inter-layer tunneling strength \cite{Anderson-1997}.

Quantum cluster calculations give very strong evidence that the
Hubbard Hamiltonian Eq.~(\ref{Eq:HtU}) is already a sufficient
minimal model to describe the main properties of the phase diagrams
of cuprates \cite{Maier-RMP-2005}. Nevertheless, these calculations
suffer from finite size effects. The simulation of this Hamiltonian
using ultra-cold fermionic atoms in an optical lattice is
straightforward and can overcome this problem for sufficiently large
lattices. It requires the loading of the atoms into the lattice and
choosing the lattice parameters such that two-dimensional layers are
created. Details on which parameter range can be simulated are
discussed in Sec.~\ref{Sec:HoppInt}.

However, the Hubbard Hamiltonian Eq.~(\ref{Eq:HtU}) does not
describe all properties encountered in a superconductor
\cite{Eskes-PRB-1996}. Other effective model Hamiltonians have been
discussed trying to incorporate such experimentally observed
effects. For example, it has been suggested
\cite{Nazarenko-PRB-1995} to introduce a next-nearest neighbor
hopping term
\begin{equation}
  \hat H_{nn} = - t' \sum_{\langle \! \langle i,j \rangle \! \rangle, \sigma}  \hat
  c^\dagger_{i,\sigma} \hat c_{j,\sigma} \,.
\end{equation}
These terms can be realized in the simulation process by using very
shallow optical lattices. In this case, however, the probability to
excite atoms into higher Bloch bands increases and off-site
interaction terms $U_{\mathrm{os}}$ are important. Also it becomes
more difficult to adjust all parameters in this Hamiltonian
independently. Therefore, in this publication, we do not discuss the
realization of this model Hamiltonian any further.

Another effective model is the so-called $t$-$J$-$U$ Hamiltonian,
which has recently been used in connection with gossamer
superconductivity \cite{Zhang-PRL-2003,Laughlin-cond-mat-2002}. It
is given by
\begin{equation} \label{Eq:Hamiltonian}
  \hat H = \hat H_{tU} + \hat H_J \,,
\end{equation}
where
\begin{equation}\label{Eq:HJ}
  \hat H_{J} = J \sum_{\langle i,j \rangle} \hat{\mathbf{S}}_i \cdot
  \hat{\mathbf{S}}_j
\end{equation}
is the spin interaction Hamiltonian of electrons in neighboring
lattice sites. The spin operator $\hat{\mathbf{S}}_j$ is defined by
its components
\begin{equation}
  \hat{\mathbf{S}}^w = \frac{1}{2} \sum_{\varsigma, \zeta = \up,\down}
   \hat c^\dagger_\varsigma \sigma^w \hat c_\zeta \,,
\end{equation}
where $\sigma^w$ are Pauli matrices, $w = x,y,z$. For very large
interaction strengths $U$ Hamiltonian Eq.~(\ref{Eq:Hamiltonian})
converges to the so-called $t$-$J$ model, where double-occupancy of
a single lattice site is excluded. This Hamiltonian has also very
often been used to describe the behavior of high-$T_c$ cuprates. In
the following sections we will investigate in detail how the
$t$-$J$-$U$ Hamiltonian can be simulated using atoms in optical
lattices. Such simulations can help to understand and test features
of the model, and also to investigate a larger parameter space which
is likely to exhibit very interesting phase diagrams.

\section{Simulation of the $t$-$J$-$U$ model with atoms in optical lattices
\label{Sec:Sim_in_latt}}

In this section we show how Hamiltonian Eq.~(\ref{Eq:Hamiltonian})
can be simulated by using ultra-cold atoms in optical lattices.
First we explain how to implement the $t$-$U$-Hamiltonian and the
spin-interaction independently from each other. Then these
Hamiltonians are combined to realize the full $t$-$J$-$U$
Hamiltonian.

\subsection{The hopping and interaction terms \label{Sec:HoppInt}}

The Hamiltonian $\hat H_{tU}$, Eq.~(\ref{Eq:HtU}), is simulated
using a three-dimensional optical lattice where hopping along the
$z$-direction is suppressed by high potential barriers
\cite{Jaksch-Ann-2005}. The lattice is filled with ultra-cold
fermionic atoms occupying the lowest motional Bloch band only and
moving in planes parallel to the $xy$-plane. We restrict our
considerations to one such 2D layer of the optical lattice. The two
spin states of the electrons are represented by two long-living
internal states of the atoms which we denote by $\sigma = \up,
\down$. The Hamiltonian which describes the dynamics of fermionic
atoms in the lattice is given by $\hat H_{t U_{s}}$,
cf.~Eq.~(\ref{Eq:HtU}), with the hopping constant $t$ and the
interaction strength $U_s$. By an appropriate choice of the lattice
constants the system parameters $t$ and $U_s$ can be tuned over a
wide range \cite{Jaksch-Ann-2005} and using a Feshbach resonance it
is possible to change the $s$-wave scattering length between the
atoms which gives additional control over the interaction strength
$U_s$.

However, we are not completely free in our choice of the lattice
depths. In order to observe the superconducting phase in high-$T_c$
cuprates the temperature has to be lower than the critical
temperature $T_c$, which is for high-temperature superconductors of
the order of $0.02 t$ \cite{Dagotto-RMP-1994}. For our system with
atoms in an optical lattice we expect the same behavior. We consider
a lattice depth of $5 E_\mathrm{R}$ with $E_\mathrm{R} = \hbar ^2 (2
\pi)^2/2 m \lambda^2$ the recoil energy, $m$ the mass of the atoms
and $\lambda$ the wave length of the laser producing the lattice
potential. In this case the hopping term $t$ is of the order of
$0.07 E_\mathrm{R}$ and thus the superconducting phase will be
observable for temperatures lower than approximately $0.0014
E_\mathrm{R}/k_\mathrm{B}$. For the fermionic species $^6$Li trapped
in a lattice with $\lambda = 670$nm this corresponds to a
temperature of 5nK. In recent experiments with $^6$Li temperatures
of about 30nK have been reported \cite{Zwierlein-Nature-2005}. The
required lower temperatures can be reached for example with phonon
side band cooling \cite{Griessner-PRA-2005, Griessner-PrivCom-2005}.

As an aside we note that in the case of bosonic atoms we have to
assume a very high interaction strength $U_{\up\up}=U_{\down \down}$
between two atoms in the same state to realize the above Hamiltonian
$\hat H_{t U_s}$ and replace anticommutator relations with
commutator relations. This interaction has to be much larger than
the inter-species interaction strength $U_{\up \down}=U_s$, the
hopping constant $t$ and the spin-spin interaction strength $J$. In
this case all states with two or more identical atoms in the same
lattice sites can be discarded. However, a direct mapping to a
fermionic system using a Jordan-Wigner transformation is only
possible in one spatial dimension \cite{Parededs-PRL-2003}. There
are proposals for similar mappings in higher dimensions which
require the addition of Majorana fermions to the system
\cite{Ball-PRL-2005,Verstraete-JSM-2005}. Although appealing from a
theoretical point of view it is not clear how to realize such
systems with atoms in optical lattices. Nevertheless, in the case of
bosons the above conditions on the interaction strength lead to a
rich phase diagram similar to models discussed in
Ref.~\cite{Duan-PRL-2003}.

\subsection{Implementation of the spin-spin interaction}

In general, the van der Waals interaction between two atoms in
neighboring lattice sites is not sufficient to realize the required
ratios of $J$, $t$, and $U$. Lowering the lattice barriers such that
there is a significant nearest neighbor interaction also causes
problems since in this case it is more likely that higher Bloch
bands are occupied. Hence the spin interaction $J$ has to be
simulated indirectly. Methods for achieving this have already been
detailed in Refs.~\cite{Sorensen-PRL-1999,
Duan-PRL-2003,Jane-QIC-2003}. These proposals require a filling
factor of very close to or exactly one atom per lattice site which
corresponds to half-filling of the electron system and thus a
Mott-insulating phase \cite{Damascelli-RMP-2003}. Therefore, these
proposals cannot be used to simulate superconductivity.

In order to circumvent this problem we extend the scheme proposed in
Ref.~\cite{Sorensen-PRL-1999}. Let us first consider the
one-dimensional spin-spin interaction in $z$-direction, i.e.,
\begin{equation}
\begin{split} \label{Eq:Spinz}
  \hat H_{zz} &= J_z \sum_{\langle i,j \rangle} \hat{S}^z_i
  \hat{S}^z_j  \\
  &= \frac{J_z}{4} \sum_{\langle i,j \rangle}
  \hat n_{\up}^i \hat n_{\up}^j + \hat n_{\down}^i \hat
  n_{\down}^j
  - \hat n_{\up}^i \hat n_{\down}^j - \hat n_{\down}^i \hat n_{\up}^j   \,.
\end{split}
\end{equation}
This type of interaction between the atoms is realized by state
selectively moving atoms. They are stored in very deep optical
lattices such that any hopping is strongly suppressed, i.e.,~$t=0$.
The atoms in, e.g., state $\ket{\up}$ are shifted and overlapped for
a certain time with their neighboring atoms to the left and to the
right in state $\ket{\down}$. The resulting atom-atom interaction
leads to the desired phase shift as described in
Ref.~\cite{Jaksch-PRL-1999}. For illustration let us assume two
lattice sites described by the Fock states
$\ket{n^1_\up,n^1_\down;n^2_\up,n^2_\down}$, where $n^j_\sigma$ is
the number of atoms in state $\ket{\sigma}$ in the $j$th lattice
site. After the above shifting procedure this state evolves to
\begin{equation}
  \ket{n^1_\up,n^1_\down;n^2_\up,n^2_\down} \longrightarrow
  \exp( \ii \chi (n^1_\down n^2_\up + n^1_\up n^2_\down))
    \ket{n^1_\up,n^1_\down;n^2_\up,n^2_\down}  \,,
\end{equation}
where $\chi$ is the phase acquired during the collision process.
This is exactly the action of the last two summands $- \hat
n_{\up}^i \hat n_{\down}^j - \hat n_{\down}^i \hat n_{\up}^j $ of
Eq.~(\ref{Eq:Spinz}).

However, this process is not yet sufficient to implement the spin
interaction in $z$-direction. In order to achieve our goal the spin
state of the atoms in every second lattice site has to be flipped,
i.e.,~the operation $V_\mathrm{fl}=\sigma^x_1\otimes \openone_2
\otimes \sigma^x_3 \otimes \openone_4 \otimes ... $ has to be
applied to the atoms. This operation requires addressing each second
lattice site which can be done by using an additional standing wave
laser field. The resulting superlattice must have twice the
wavelength of the original lattice and its intensity minima need to
coincide with every second lattice site of the trapping potential
\cite{footnote2}. Thus the energy levels of every second atom are
AC-Stark shifted out of resonance such that a microwave or
laser-field driving the transition $\ket{\up} \leftrightarrow
\ket{\down}$ realizes $V_\mathrm{fl}$.

By repeating this shifting process after the spin flip operation
$V_\mathrm{fl}$ and flipping the atoms back the state evolves
according to
\begin{equation}
  \ket{n^1_\up,n^1_\down;n^2_\up,n^2_\down} \longrightarrow
  \exp( -\ii \chi (n^1_\down n^2_\down + n^1_\up n^2_\up))
    \ket{n^1_\up,n^1_\down;n^2_\up,n^2_\down}  \,,
\end{equation}
where the collision time has to be chosen such that the acquired
phase is equal to $-\chi$. The whole process induces dynamics
according to Hamiltonian Eq.~(\ref{Eq:Spinz}), where $\chi = J_z
\tau/4\hbar$ and $\tau$ is the time for which $\hat H_{zz}$ is
applied (details on calculating the phases can be found in Appendix
\ref{app-phase}).

The creation of such spin-dependent phases has already been
demonstrated experimentally. In Ref.~\cite{Mandel-Nature-2003} the
authors used Rb atoms in an optical lattice with $V_0 = 34
E_\mathrm{R}$. Excitations of atoms into higher Bloch bands were
avoided by using an appropriate time of $\tau_s=40\mu\mathrm{s}$ to
shift the atoms spin-dependently into the neighboring lattice sites
\cite{Mandel-PRL-2003}. After holding the atoms in the shifted
position for $\tau_h=450\mu\mathrm{s}$ and shifting them back again
collisional phases of $\chi \approx 2 \pi$ were achieved.

With being able to implement Hamiltonian Eq.~(\ref{Eq:Spinz}) it is
also possible to realize the Hamiltonians
\begin{gather}
  \hat H_{xx} = J_x \sum_{\langle i,j \rangle} \hat{S}^x_i
  \hat{S}^x_j \,, \\
  \hat H_{yy} = J_y \sum_{\langle i,j \rangle} \hat{S}^y_i
  \hat{S}^y_j \,.
\end{gather}
We first observe that with suitable laser pulses it is possible to
implement the following rotations on single atoms
\begin{gather} \label{Eq:Vy}
  V_y = \exp\left(   i \frac{\pi}{4} \sigma^y \right) \,, \\
  V_x = \exp\left( - i \frac{\pi}{4} \sigma^x \right) \,.
  \label{Eq:Vx}
\end{gather}
By simultaneously applying one of these rotations on all atoms,
implementing Hamiltonian Eq.~(\ref{Eq:Spinz}) and applying the
Hermitian conjugate of the same rotation we get the time evolutions
\cite{Jane-QIC-2003,Sorensen-PRL-1999,Jaksch-PRA-2002}
\begin{gather}
  V_y^\dagger \exp\left( - \ii \hat H_{zz} \frac{\tau}{\hbar} \right) V_y
  = \exp\left( - \ii \hat H_{xx} \frac{\tau}{\hbar} \right) \,, \\
  V_x^\dagger \exp\left( - \ii \hat H_{zz} \frac{\tau}{\hbar} \right) V_x
  = \exp\left( - \ii \hat H_{yy} \frac{\tau}{\hbar} \right) \,.
\end{gather}
The interaction strengths $J_x$, $J_y$, and $J_z$ can be tuned
independently from each other by the choice of the collisional phase
$\chi$ during the simulation of the respective Hamiltonian.

\subsection{Combining the Hamiltonians}

We simulate Hamiltonian Eq.~(\ref{Eq:Hamiltonian}) using the well
known Trotter-Suzuki expansion \cite{Trotter-PAMS-1959,
Suzuki-PLA-1990, Suzuki-JMP-1991, Suzuki-JMP-1993, Suzuki-CMP-1994,
Jane-QIC-2003} described in detail in Appendix \ref{app-trotter}. In
this approach the different parts of the Hamiltonian are simulated
for a small time $\tau$ separately. For instance, in first order,
the decomposition of the $t$-$J$-$U$ Hamiltonian reads
\begin{equation}
\begin{split}
  &\exp\left( \frac{-\ii \hat H  \,\tau}{\hbar} \right) =
   \exp\left( \frac{-\ii \hat H_{zz} \,\tau}{\hbar} \right)
   V_x^\dagger\exp\left( \frac{-\ii \hat H_{zz} \,\tau}{\hbar} \right)V_x\\
   &\quad \times
   V_y^\dagger\exp\left( \frac{-\ii \hat H_{zz}\, \tau}{\hbar} \right)V_y
   \exp\left( \frac{-\ii \hat H_{tU_s}\, \tau}{\hbar} \right)
    + \mathcal{O}(\tau^2)\,.
\end{split}
\end{equation}
The last part $\exp\left( -\ii \hat H_{tU_s}\, \tau/\hbar \right)$
of this time evolution is simulated by choosing a suitable lattice
depth such that the required values of $t$ and $U_s$ are realized.
After waiting a time $\tau$ \cite{footnote3} the lattice has to be
ramped up avoiding any excitation into higher Bloch bands. Then the
shifting and flipping processes $V_\mathrm{fl}$, as described in the
previous section, have to be applied in order to simulate the time
evolution of $\hat H_{zz}$, $\hat H_{yy}$, and $\hat H_{xx}$. The
latter two require the implementation of the operations $V_x$ and
$V_y$, cf.~Eqs.~(\ref{Eq:Vy}) and (\ref{Eq:Vx}). By repeating all of
these simulation steps $m$ times a time $m\tau$ is simulated with an
error $\propto \tau^2 m$. Better accuracies can be achieved by using
higher order approximations, see Appendix \ref{app-trotter}.

During the simulation of the spin-spin interaction term two atoms
may occupy the same lattice site $j$. This gives rise to an
additional term not considered so far. We illustrate this by
considering shifts in $x$-direction during the simulation of the
spin interaction. The atoms will acquire an additional phase
(compare Appendix \ref{app-phase})
\begin{equation}
  \phi = K  \int_{-\tau_0}^{\tau_0} \! \diff \tau \, \exp\left(
  -\frac{[x_\up^j(\tau) - x_\down^{j}(\tau)]^2}{2 x_0^2}\right)\,,
\end{equation}
with $K$ given in Eq.~(\ref{Eq:K}). This phase corresponds to an
(additional) effective interaction term
\begin{equation}
  \hat H_{U}^\mathrm{eff} =  U_\mathrm{eff} \sum_j \hat n_{j,\up} \hat
  n_{j,\down}\,,
\end{equation}
where the interaction constant is given by $U_\mathrm{eff} = \hbar
\phi /\tau$. Note that this effective interaction does not diverge
for $\tau \to 0$ as in this case also $\phi$ goes to zero. This
interaction provides an additional opportunity for tuning the
simulated interaction constant $U$. The total interaction for small
$U_\mathrm{eff} \ll 2 \pi \hbar / \tau$ is given by $U = U_s + 3d
U_\mathrm{eff}$ with $d$ the number of spatial dimensions in which
the spin-spin interaction is simulated. Here the factor of $3$ in
front of the dimension number $d$ arises because the phase $\phi$
occurs in the simulations of $\hat H_{xx}$, $\hat H_{yy}$, and $\hat
H_{zz}$ which are all simulated for the same time $\tau$. By an
appropriate choice of $U_s$ and $\phi$ it is possible to simulate
effective interactions which are attractive. For this purpose, the
phase $\phi$ has to be chosen in such a way that it can be written
as $\phi = 2 \pi + U'_\mathrm{eff}\tau/\hbar$ with a small negative
$U'_\mathrm{eff}$ and $2 \pi \gg |U'_\mathrm{eff}|\tau/\hbar$. The
total simulated interaction is then given by $U = U_s + 3d
U'_\mathrm{eff}$. For not too large $U_s$ this leads to a negative
$U$ \cite{foot1}.

\begin{figure}
  \includegraphics[width = 7 cm]{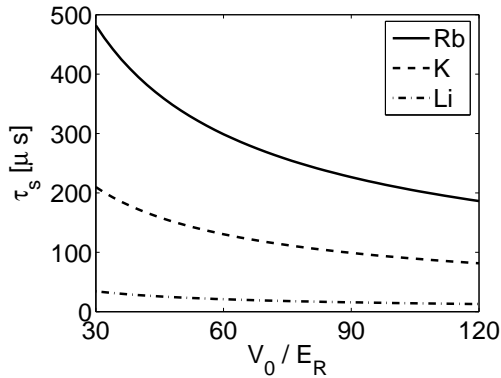}
  \caption{Lower bounds for the calculated shifting and holding
  times $\tau_{sh}$ (cf.
  Eq.~(\ref{Eq:ShiftTimes})) for lattices with $^{87}\mathrm{Rb}$ atoms
  ($\lambda = 826 \mathrm{nm}$, $a_s = 5.1 \mathrm{nm}$),
  $^{40}\mathrm{K}$ atoms ($\lambda = 826 \mathrm{nm}$, $a_s = 5.5 \mathrm{nm}$)
  and $^{6}\mathrm{Li}$ atoms
  ($\lambda = 670 \mathrm{nm}$, $a_s = 2.4 \mathrm{nm}$). The values
  given in brackets are typical values used for the calculation of
  $\tau_{sh}$.
  \label{Fig:ShiftTimes}  }
\end{figure}

The total time for which the system can be simulated is restricted
by the life- and decoherence-time of the ultra-cold atoms in the
optical lattice. This is typically on the order of one second
\cite{Greiner-Nature-2002}. Hence we have to estimate the time which
is needed for simulating one time step of Hamiltonian
Eq.~(\ref{Eq:Hamiltonian}). During the shifting process heating of
the atoms has to be avoided. Since the lattice is required to be
very deep during the atoms are shifted hopping is strongly
suppressed. This leads to a flat Bloch band and thus no excitation
can take place within the band during the shift. Any heating thus
means that atoms are excited into higher Bloch bands. In order to
avoid such excitations we require for each shift of atoms in,
e.g.,~state $\ket{\up}$ into the neighboring lattice site a time
larger than the inverse of the lattice site trapping frequency
$\omega_t = \sqrt{V_0 / E_\mathrm{R}} \, \hbar (2
\pi)^2/m\lambda^2$, as already demonstrated experimentally in
Refs.~\cite{Mandel-Nature-2003,Mandel-PRL-2003}. Furthermore, the
atoms have to be held in the shifted position for a certain time in
order to achieve the desired phase. This phase is either $\chi$ or
$2\pi - \chi$. Thus the average time for one shifting and holding
process is given by
\begin{equation}\label{Eq:ShiftTimes}
  \tau_{sh} > 2 \pi \left( \frac{4}{\omega_t} + \frac{1}{K}\right)\,.
\end{equation}
The minimal shifting times for varying lattice depths are shown in
Fig.~\ref{Fig:ShiftTimes} and are of the order of a few hundred
microseconds for Rb or a few tens of microseconds for Li. To
simulate the time evolution of $\hat H_{zz}$ these shifts have to be
repeated $2d$ times.

The constraints for the ramping time between the application of the
$t$-$U$-Hamiltonian and the spin-spin interaction are twofold: The
ramping has to be adiabatic on a time scale given by $1/\omega_t$ to
avoid excitations into higher Bloch bands and it has to be rapid
compared to $\hbar \pi/2t$. This ensures that the system does not
evolve for too long with a hopping term $t$ that is different from
the desired one or even follows the change in $t$ adiabatically in
contrast to the sudden change required by the Trotter-Suzuki
expansion. The first constraint has been experimentally tested for
Rb atoms \cite{Greiner-Nature-2002b} and was found to be fulfilled
for ramping times longer than $50\mu\mathrm{s}$ in accordance with
theoretical calculations \cite{Clark-PRA-2004}. For typical lattice
parameters the hopping time is given by $\hbar \pi/2t =
1\mathrm{ms}$ and is thus more than one order of magnitude larger
than the minimal ramping time required to avoid excitations into
higher Bloch bands. Therefore, both constraints can be fulfilled at
the same time. Furthermore, the ramping process conserves
quasi-momentum, and excitations that do not change the
quasi-momentum involve at least two hopping processes. Because the
lattice is ramped on a time scale short compared to $\hbar \pi/2t$
excitations within the lowest Bloch band are, therefore, also
strongly suppressed. Together with the time needed to simulate the
$t$-$U$-Hamiltonian it is thus feasible to simulate several hundreds
of Trotter steps $m$ within the lifetime of the atoms in the
lattice.

We finally remark that this procedure for simulating the $t$-$J$-$U$
Hamiltonian is compatible with the additional simulation of a
magnetic field. Various methods for creating effective magnetic
fields at the same time as the hopping and interaction terms in
$\hat H_{tU}$ were recently proposed \cite{Jaksch-NJP-2003,
Juzeliunas-PRL-2004, Juzeliunas-PRA-2005, Juzeliunas-PRA-2006,
Mueller-PRA-2004, Sorensen-PRL-2005}. Thus it is possible to extend
our setup for studying the $t$-$J$-$U$ Hamiltonian in external
magnetic fields.

\subsection{Measuring the properties of the atom gas}

The properties of the time-evolved state can be probed by
measurements of first- and second-order correlation functions as
proposed in Ref.~\cite{Altman-PRA-2004}. For this purpose a multiple
matter-wave version of the Hanbury Brown and Twiss experiment
\cite{Hanbury-Nature-1956} has to be realized. The atoms are
released from their trapping potential and imaged after a certain
time of flight. Depending on the phase of the atomic gas these
images reveal interference patterns; second-order correlation
functions give additional information on the phase and the
underlying structure of the lattice. Such measurements have already
been performed for bosonic atoms in the Mott-insulating
phase~\cite{Greiner-Nature-2002,Foelling-Nature-2005} and agree very
well with the theoretical predictions~\cite{Oosten-PRA-2001}. With
these techniques it will be possible to check the phase diagrams of
the simulated systems and compare them to calculated or measured
phase diagrams for different high-temperature superconductors (see,
e.g.,~\cite{Dagotto-RMP-1994, Damascelli-RMP-2003, Wen-PRL-1996}).

\section{Numerical simulations \label{Sec:Numerics}}

As already mentioned in the introduction the simulation of
Hamiltonian Eq.~(\ref{Eq:Hamiltonian}) is a very hard computational
task. Even for a few lattice sites an exact simulation of the system
is no longer feasible and one has to use approximations such as
mean-field theories or Quantum Monte Carlo calculations. Since we
want to compare the results of the simulation of Hamiltonian
Eq.~(\ref{Eq:Hamiltonian}) with exact results we have thus
restricted ourselves to the one-dimensional case with a few lattice
sites only.

The anti-fidelity of the simulated state compared to a state which
is time-evolved using the full Hamiltonian
Eq.~(\ref{Eq:Hamiltonian}) is calculated as follows. Let $\teopnull
(\tau) = \exp(- \ii \hat H \tau /\hbar)$ be the full time evolution
operator, $\teop (\tau)$ the simulated time evolution,
$\ket{\psi}_\mathrm{in}$ an input state and $\ket{\psi} =  \teop
(\tau)\ket{\psi}_\mathrm{in}$, $\ket{\psi_0} = \teopnull
(\tau)\ket{\psi}_\mathrm{in}$. Due to the Cauchy-Schwarz inequality
and the properties of the matrix norm we get
\begin{equation}
\begin{split}\label{Eq:App1}
  || \teop (\tau) - \teopnull(\tau)||^2 \geq & | ( \bra{\psi} - \bra{\psi_0})
  (\ket{\psi} - \ket{\psi_0})|\\
   & = |2 - 2 \Re(\left\langle \psi |
  \psi_0 \right\rangle ) |\,.
\end{split}
\end{equation}
Since $|\left\langle \psi | \psi_0 \right\rangle| \leq 1$ the rhs of
Eq.~(\ref{Eq:App1}) is always positive even without the modulus and
we can rearrange
\begin{equation}
  |\left\langle \psi | \psi_0 \right\rangle| \geq 1 - \frac{1}{2}
  || \teop (\tau) - \teopnull (\tau)||^2 \,.
\end{equation}
For $ ||\teop (\tau) - \teopnull (\tau)||^2 \leq 2$ this yields an
upper bound for the anti-fidelity $\mathcal{F}$
\begin{equation}
  {\cal F} = 1 - |\left\langle \psi | \psi_0 \right\rangle|^2 \leq ||\teop (\tau) -
  \teopnull (\tau)||^2 \,,
\end{equation}
which can be easily calculated from the time evolution operators.

\begin{figure}[t]
  \includegraphics[width = 7 cm]{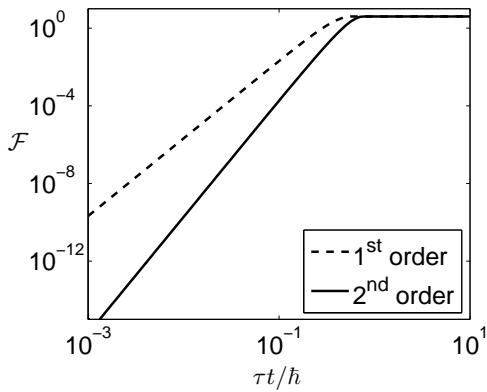}

  ~\qquad \hfill \raisebox{1.8ex}[-1.8ex]{$\tau t/\hbar$}  \hfill~
  \caption{Simulation of Hamiltonian Eq.~(\ref{Eq:Hamiltonian})
  using first- and second-order Trotter expansions,
  cf.~Eqs.~(\ref{EQ:simO2}) and (\ref{Eq:simO3}). Upper bounds
  for the anti-fidelity $\mathcal{F}$ are shown.
  The number of lattice sites is $M=5$,
  the parameters for the simulation are $J = 0.3 t$,
  $U_s = 5 t$, $U_\mathrm{eff}' = -2 t$, $m=1$ leading to $U=-t$.
  \label{Fig:TimeSim}  }
\end{figure}

We have calculated this anti-fidelity $\mathcal F$ for several
cases. First, we investigate a parameter set which gives a negative
(attractive) interaction strength $U$ and compare this to the exact
time evolution according to Eq.~(\ref{Eq:Hamiltonian}). Upper bounds
for the corresponding anti-fidelity $\mathcal F$ are shown in
Fig.~\ref{Fig:TimeSim}. It has been assumed that no errors such as
unprecise creation of the phase shifts $\chi$ or excitation of
higher Bloch bands occur during the shifting process. For small
times the slopes of the curves for the first and second order
approximations agree very well with the predictions from first and
second order expansions, cf.~Eqs.~(\ref{EQ:simO2}) and
(\ref{Eq:simO3}). Note that the slopes shown are increased by a
factor of two because we take the square of the matrix norm in order
to calculate the anti-fidelity $\mathcal F$. Figure
\ref{Fig:TimeSim} shows that for small times the simulation of the
time evolution of Hamiltonian Eq.~(\ref{Eq:Hamiltonian}) agrees very
well with the full time evolution but the desired simulation times
of order $100 \hbar/t$ can only be reliably achieved by repeating
this simulation process.

\begin{figure}
  \includegraphics[width = 7 cm]{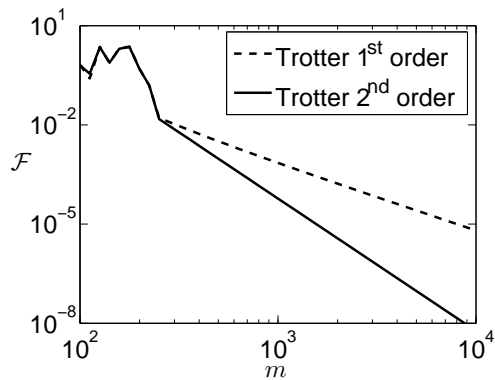}
  \caption{Upper bounds for the anti-fidelity $\mathcal{F}$ for first- and
  second-order Trotter expansions, cf.~Eq.~(\ref{Eq:Trotter}). The simulated time is fixed to
  $\tau = 100 \hbar/t $, the number of lattice sites is $M=5$.
  We have used $J = 0.3 t$,  $U_s = 10t$, and $\phi = 0$.
  \label{Fig:TimeTrotter}  }
\end{figure}
The dependence of the upper bounds of the anti-fidelity $\mathcal F$
on the number $m$ of Trotter steps for a fixed time $\tau = 100
\hbar /t$ is shown in Fig.~\ref{Fig:TimeTrotter} for values
corresponding to typical parameters in high-temperature
superconductivity \cite{Anisimov-PRB-2002}. As expected the fidelity
gets better for an increasing number of steps. In second order
approximation for $m= 500$ steps the anti-fidelity is already
smaller than $10^{-3}$. For the same accuracy using the first-order
approximation $m = 900$ Trotter steps have to be used. This means in
first-order approximation the lattice has to be ramped up and down
900 times and $\hat H_{zz}$ has to be simulated 2700 times with 1800
applications of $V_{x,y}$. In second-order each single step is more
complicated, but by choosing the simulation process in a judicious
way only 500 rampings of the lattice are required. By optimizing the
process (cf.~Appendix~\ref{app-trotter}) only 2500 simulations of
$\hat H_{zz}$ are required with 1500 applications of $V_{x,y}$.

\begin{figure}
  \includegraphics[width = 7 cm]{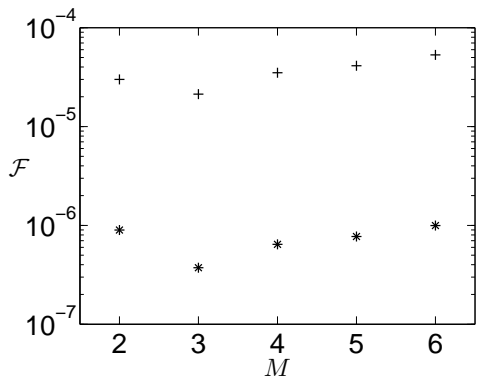}
  \caption{Upper bounds for the anti-fidelity $\mathcal{F}$ vs number
  of lattice sites $M$ for first- (pluses) and second-order
  (stars) Trotter expansion.  The simulated time was $\tau = 0.01 \hbar /t$
  and we have chosen $J=0.3 t$, $U_s = 10t$ and $\phi = 0$.
  \label{Fig:Mdependence}  }
\end{figure}

To get an impression on how the anti-fidelity $\mathcal F$ scales
with the number of lattice sites $M$ we calculated it for different
values of lattice sites $M$ in a 1D lattice. The results are shown
in Fig.~\ref{Fig:Mdependence} for a simulated time of $\tau = 0.01
\hbar/t$. The results indicate that the anti-fidelity $\mathcal F$
does not increase quickly with the number of lattice sites.
Therefore, it should be possible to experimentally simulate systems
of realistic size.

Since the time evolution of the full Hamiltonian
Eq.~(\ref{Eq:Hamiltonian}) is only simulated approximately in our
scheme, the state vector of the atoms after the time evolution will
also only approximate the real state. However, we expect errors due
to this effect to be small, since the necessary measurements to
distinguish the phases involve only first- and second-order
correlation functions. The influence of small imperfections of the
simulated Hamiltonian on these functions will in general be smaller
than the influence on the state vector estimated by the upper bounds
of the anti-fidelity $\mathcal{F}$. Thus the deviations of the
simulated phase diagrams is expected to be smaller than implied by
the anti-fidelities.

\section{Conclusion \label{Sec:Concl}}

In the present paper we have shown that it is feasible to simulate
different model Hamiltonians which are used to describe
high-temperature superconductive cuprates. The simulation of the
most minimal model, the Hubbard model Eq.~(\ref{Eq:HtU}), can
already be expected to deepen our understanding of the phase
diagrams encountered in the cuprates. Since this model does not
describe all experimentally measured effects and to enlarge the
accessible parameter space which can be investigated using atoms in
optical lattices we have especially discussed how to simulate the
$t$-$J$-$U$ Hamiltonian Eq.~(\ref{Eq:Hamiltonian}), which requires
the inclusion of an additional spin-spin interaction term $J$.

In order to make the simulation possible we have extended earlier
proposals to create an effective spin interaction between atoms to
the case where the lattice is not fully occupied. We found that near
future technology will allow to simulate Hamiltonian
Eq.~(\ref{Eq:Hamiltonian}) for times which are longer than the
typical time scales associated with the dynamics of the $t$-$J$-$U$
Hamiltonian. Hence the properties of the Hamiltonian can be made
visible using our simulation method. We also showed that by an
appropriate choice of the lattice parameters and shifting times it
is possible to create attractive on-site interactions $U$. Also, our
method is compatible with the simulation of magnetic fields in the
lattice.

Furthermore, we simulated the simulation process numerically for
small systems and compared the results with exact calculations of
the time evolution. For these simulations the results show very good
agreement between the simulated and the real time evolution.
Anti-fidelities ${\cal F} < 10^{-3}$ can be achieved. Because of
these results and of the experimental progress we are confident that
the proposed method will help to deepen our knowledge and
understanding of the $t$-$J$-$U$ Hamiltonian model.

\acknowledgments{A.K.~thanks Stephen R.~Clark for useful discussions
 and acknowledges financial support from the Keble
 Association. This work was supported by the EPSRC (UK) through
 QIP IRC (GR/S82176/01) and project EP/C51933/1. The research
 was also supported by the EU through the STREP project OLAQUI
 (http://olaqui.df.unipi.it/).
}

\appendix
\section{Calculation of the collisional phases}
\label{app-phase}

The exact phase shift due to the shift and interaction process can
be calculated analytically~\cite{Jaksch-PRL-1999}. If we assume deep
lattice potentials of $V_{x,y,z}$, we can approximate the wave
functions of the atoms located in one lattice site by Gaussians. For
illustration we assume an atom of state $\ket{\up}$ in lattice site
$j$ and one of state $\ket{\down}$ in lattice site $j+1$. The shift
shall occur in $x$-direction. Let $x^j_\up(\tau)$,
$x^{j+1}_\down(\tau)$ denote the time-dependent $x$-coordinates of
the atoms. Because of the collision process the state of the two
atoms will evolve to
\begin{equation}
  \ket{1,0;0,1} \to \eh^{- \ii \chi}\ket{1,0;0,1}
\end{equation}
with the collisional phase given by
\begin{equation}
  \chi = K \int_{-\tau_0}^{\tau_0} \! \diff \tau\, \exp\left(
  -\frac{[x_\up^j(\tau) - x_\down^{j+1}(\tau )]^2}{2
  x_0^2}\right)\,.
\end{equation}
Here we have defined
\begin{equation}\label{Eq:K}
  K = \frac{4 \pi a_s \hbar}{m} \frac{(\sqrt{2
  \pi})^3}{\lambda^3} \left(\frac{V_0}{E_\mathrm{R}}
  \right)^\frac{3}{4} \,,
\end{equation}
where $a_s$ is the $s$-wave scattering length between the two atoms
in states $\ket{\up}$ and $\ket{\down}$, $V_0$ is the depth of the
lattice, $x_0 = \sqrt[4]{E_\mathrm{R}/V_0}  \lambda/2 \pi$ and
$E_\mathrm{R}= \hbar^2 (2 \pi)^2/m \lambda^2$ is the recoil energy,
$m$ the mass of the atoms and $\lambda$ the wavelength of the lasers
creating the optical lattice. The time $\tau_0$ is chosen such that
the entire shift process is contained in the time interval from
$-\tau_0$ to $\tau_0$. For typical experimental
parameters~\cite{Mandel-Nature-2003} of $V_0 = 34 E_\mathrm{R}$ and
$^{87}\mathrm{Rb}$ atoms with $a_s = 5.1$nm the prefactor in front
of the integral has a value of roughly $K =
19\mathrm{rad}/\mathrm{ms}$ such that after a collisional time of a
few hundred microseconds phase shifts of the order of $2\pi$ can be
achieved. For Li atoms the prefactor has a value of the order of $K
= 230\mathrm{rad}/\mathrm{ms}$ and hence only roughly a tenth of the
time is necessary to achieve similar phase shifts. The exact phase
shifts depend on the details of the shifting process itself,
i.e.,~on the exact form of the functions $x_\sigma^j(\tau)$.

\section{The Trotter-Suzuki expansion}
\label{app-trotter}

For an arbitrary Hamiltonian $\hat H_{\mathrm{arb}} = \sum_{j=1}^f
\hat H_j$ the Trotter-Suzuki expansion \cite{Trotter-PAMS-1959,
Suzuki-PLA-1990, Suzuki-JMP-1991, Suzuki-JMP-1993, Suzuki-CMP-1994,
Jane-QIC-2003} is found by defining
\begin{gather} \label{EQ:simO2}
  Q_1(x) = \prod_{j = 1}^f \exp( x \hat H_j) \,, \\
  Q_2(x) = \eh^{\hat H_1 x/2} ... \eh^{\hat H_{f-1} x/2} \eh^{\hat
  H_f x} \eh^{\hat H_{f-1} x/2} ... \eh^{\hat H_1 x/2} \,,
  \label{Eq:simO3}
\end{gather}
where $x = - i \tau /\hbar$ and $\tau$ is the simulation time. The
operators $Q_1(x)$ and $Q_2(x)$ are approximations to the time
evolution operator of $\hat H_{\mathrm{arb}}$ with
\begin{equation}
  \exp( \hat H_{\mathrm{arb}} x) = Q_1(x) + \mathcal{O}(|x|^2) =
  Q_2(x) + \mathcal{O} (|x|^3) \,.
\end{equation}
Higher order approximations to the time evolution of $\hat
H_{\mathrm{arb}}$ exist, but they either require the simulation of
negative time which is difficult for the hopping Hamiltonian $\hat
H_{tU} $, or they include terms which make the simulation unstable
\cite{Suzuki-PLA-1990}. Therefore, we simulate longer times using a
generalized version of the Trotter formula \cite{Suzuki-CMP-1994}
with $m$ simulation steps, reading
\begin{equation}\label{Eq:Trotter}
  \exp( \hat H_{\mathrm{arb}} x) = \left[ Q_j\left( \frac{x}{m}
  \right) \right]^m + \mathcal{O} \left( \frac{|x|^{j+1}}{m^j} \right) \,.
\end{equation}
When using the second-order approximation Eq.~(\ref{Eq:simO3}) in
this formula the experimental procedure can be optimized by
combining subsequent simulations of $\hat H_1$ since $\exp(\hat H_1
x/2) \exp(\hat H_1 x/2) = \exp(\hat H_1 x)$.

\bibliography{Simulator}

\begin{thebibliography}{54}
\bibitem[{foo({\natexlab{b}})}]{footnote3}
\bibinfo{note}{Alternatively, if the real values $t_r$ and $U_r$ implemented in
  the lattice are different from the desired values $t$ and $U$, it is also
  possible to let the system evolve for a real time $\tau_r$ such that $\tau t
  = \tau_r t_r$ and $\tau U = \tau_r U_r$.}
\bibitem[{foo({\natexlab{c}})}]{foot1}
\bibinfo{note}{This method can also be used to simulate negative interaction
  strengths in Hubbard models in optical lattices. Properly timed periodic
  raising and lowering of the potential barriers to switch off the hopping
  terms leads to effective negative onsite interaction strengths.}
\end{thebibliography}

\end{document}